\pdfoutput=1
\documentclass[letterpaper, 10 pt, conference]{ieeeconf}
\IEEEoverridecommandlockouts
\usepackage{cite}
\usepackage{float}
\usepackage{stfloats}
\usepackage{amsmath,amssymb}
\usepackage{graphicx}
\usepackage{booktabs}
\usepackage{xcolor}
\usepackage{url}
\usepackage{listings}
\usepackage{siunitx}
\usepackage{tikz}
\usetikzlibrary{positioning,arrows.meta,calc}
\usepackage{pgfplots}
\usepgfplotslibrary{groupplots}
\definecolor{c-scaly}{HTML}{009E73}
\definecolor{c-sx}{HTML}{D55E00}
\definecolor{c-mx}{HTML}{0072B2}
\definecolor{c-callmx}{HTML}{E69F00}
\definecolor{c-mapsx}{HTML}{CC79A7}
\definecolor{c-gemm}{HTML}{56B4E9}
\definecolor{c-casadi}{HTML}{4A3AA8}

\pgfplotsset{
  compat=1.18,
  bench/.style={
    unbounded coords=jump,
    tick label style={font=\scriptsize},
    label style={font=\footnotesize},
    title style={font=\footnotesize, yshift=-1ex},
    every axis plot/.append style={line width=0.8pt, mark size=1.8pt},
    log ticks with fixed point,
    ymajorgrids, grid style={gray!25},
  },
  bk/scaly/.style={c-scaly, mark=*, line width=1.4pt},
  bk/casadi_sx/.style={c-sx, mark=square*},
  bk/mxdash/.style={densely dashed, every mark/.append style={solid}},
  bk/casadi_mx/.style={c-mx, mark=triangle*, bk/mxdash},
  bk/casadi_call_mx/.style={c-callmx, mark=diamond*, bk/mxdash},
  bk/casadi_map_sx/.style={c-mapsx, mark=triangle*, every mark/.append style={rotate=180}},
  bk/casadi_mx_gemm/.style={c-gemm, mark=x, mark size=2.3pt, bk/mxdash},
  bk/best/.style={black!18, line width=3pt, no marks},
}

\newcommand{\benchseries}[2]{%
  \addplot[bk/best, forget plot] table[x=size, y=#2_best, col sep=comma] {#1};%
  \pgfplotsinvokeforeach{casadi_sx,casadi_mx,casadi_call_mx,casadi_map_sx,casadi_mx_gemm,scaly}{%
    \addplot[bk/##1, forget plot] table[x=size, y=#2_##1, col sep=comma] {#1};%
  }%
}

\newcommand{\timeouts}[1]{%
  \pgfplotsinvokeforeach{casadi_sx,casadi_mx,casadi_call_mx,casadi_map_sx,casadi_mx_gemm}{%
    \addplot[bk/##1, only marks, every mark/.append style={fill=white}, forget plot] table[x=size, y=to_##1, col sep=comma] {#1};%
  }%
}

\newcommand{\benchratio}[4]{%
  \pgfplotstableread[col sep=comma]{#1}\ratiotable
  \pgfplotstablegetelem{#3}{size}\of\ratiotable\let\ratiox\pgfplotsretval
  \pgfplotstablegetelem{#3}{#2_scaly}\of\ratiotable\let\ratioa\pgfplotsretval
  \pgfplotstablegetelem{#3}{#2_best}\of\ratiotable\let\ratiob\pgfplotsretval
  \pgfplotstablegetelem{#3}{#2_ratio}\of\ratiotable\let\ratioy\pgfplotsretval
  \edef\ratiodraw{\noexpand\draw[{Latex[length=0.9mm, width=1mm]}-{Latex[length=0.9mm, width=1mm]}, shorten <=0.5pt, shorten >=0.5pt]
    (axis cs:\ratiox,\ratioa) -- node[#4, font=\noexpand\scriptsize, inner sep=1pt, fill=white]
    {$\noexpand\pgfmathprintnumber[fixed, precision=1]{\ratioy}\noexpand\times$} (axis cs:\ratiox,\ratiob);}%
  \ratiodraw
}

\makeatletter
\let\NAT@parse\undefined
\makeatother
\usepackage[hidelinks]{hyperref}
\newcommand{\scaly}{Scaly}
\newcommand{\casadi}{CasADi}

\title{\LARGE \bf Scaly: an Optimizing Compiler for Optimal Control}

\author{Tudor A. Oancea$^{1}$ and Colin N. Jones$^{1}$%
\thanks{This work was supported by the Swiss National Science Foundation through the NCCR Automation (grant agreement 51NF40\_180545).}%
\thanks{$^{1}$Automatic Control Laboratory (LA3), EPFL, Lausanne, Switzerland. {\tt\small \{tudor.oancea, colin.jones\}@epfl.ch}}%
}

\begin{document}
\bstctlcite{IEEEexample:BSTcontrol}
\maketitle
\thispagestyle{plain}
\pagestyle{plain}

\begin{abstract}
We present \scaly{}, an open-source optimizing compiler for optimal control, written in Python, that generates fast C code.
\scaly{} provides a modeling language embedded in Python for the parametric optimization problems that arise in model predictive control and moving horizon estimation, and computes the sparse derivatives that standard optimization solvers need.
\scaly{} accepts general nonlinear programs but applies optimizations specific to optimal control, which preserve the repeated structure of a problem, such as the stages of a horizon, through differentiation down to the generated loops.
The generated C depends only on the C math library and never allocates memory dynamically.
On three benchmark problems, \scaly{} evaluates the Hessian of the Lagrangian up to 26 times faster than \casadi{}, with smaller binaries and build times that stay nearly constant with the horizon.
With the same solver, \scaly{} shortens the solver time per control step on every benchmark, by a factor of up to 9.
\end{abstract}

\section{Introduction}\label{sec:intro}

Nonlinear model predictive control (NMPC)~\cite{rawlings2017mpc} computes the control input at each sampling time by solving a finite-horizon optimal control problem (OCP) online, starting from the current state estimate.
Besides solving the linear systems that arise in its iterations, the solver spends a large share of its time evaluating the cost, the constraints, and their first- and second-order derivatives, which we call \emph{function evaluation}.
The code of these functions is often generated by a symbolic modeling tool embedded in a high-level language such as Python or Matlab, which also obtains the derivatives by automatic differentiation (AD)~\cite{griewank2008evaluating}.
Since dimensions and sparsity patterns stay fixed across the successive runs of the controller, the generated C code can be specialized, requiring neither dynamic memory nor external libraries, and can be used from prototyping on a workstation to deployment on an embedded target.
Early code generators specialized an entire solver to a family of problems with fixed dimensions, such as ACADO~\cite{houska2011acado} for NMPC and CVXGEN~\cite{mattingley2012cvxgen} and FORCES~\cite{domahidi2012forces} for convex problems.
\casadi{}~\cite{andersson2019casadi}, the de facto standard today, generates the problem functions separately from the solver and underlies much of the open-source NMPC software~\cite{verschueren2022acados}.

Recent NMPC formulations use longer horizons and larger systems and include learned components such as neural dynamics models~\cite{salzmann2023rtn, waibel2026npmpc}.
These formulations combine the sparse, repeated structure of an OCP with dense computations, such as the matrix-vector products of a neural network layer evaluated at every stage.
A code generator for these controllers therefore has to produce fast code for both small first-principles problems and large problems with dense learned parts.

Existing compilers for optimal control, such as \casadi{}~\cite{andersson2019casadi} and ACADO~\cite{houska2011acado}, analyze sparsity mainly to compute sparse derivatives, perform few optimizations after differentiation, and translate their representation of the problem into C code in essentially one step.
The generated code therefore either expands every operation into scalar arithmetic, which simplifies away the zeros of a derivative but repeats the code of every stage, so that the C source grows with the horizon until compiling it takes a great deal of time, or keeps matrix operations and loops, which stays compact but forgoes these simplifications (Section~\ref{sec:fast-code}).
For example, the scalar code that \casadi{} generates for the Hessian of an MPC with a first-principles vehicle model reaches several megabytes at long horizons, and the code of a safety filter with neural dynamics for many cars does not compile within minutes in any form (Section~\ref{sec:experiments}).

Optimization problems in control often apply the same function many times to subsets of their variables, at every stage of the horizon of an OCP or for every pair of agents in multi-agent constraints.
An optimizing compiler, which keeps the problem in its own internal form and transforms it in several steps before writing C, can automatically preserve this repetition through differentiation down to the generated code, so that the code of an OCP does not grow with its horizon.
Machine-learning compilers, such as those of JAX~\cite{jax} and PyTorch~\cite{ansel2024pytorch2}, already transform programs in this way, but they do not exploit the fixed sparsity patterns of derivatives and do not generate dependency-free C code (Section~\ref{sec:background}).

We present \scaly{}, an optimizing compiler for optimal control that brings these program transformations together with automatic sparse differentiation.
Its contributions are:
\begin{itemize}
  \item a compiler architecture for OCP code generation with two intermediate representations, one for the mathematics of a function and one for its execution, so that the compiler can transform the algebra of a function directly and reuse memory buffers efficiently (Section~\ref{sec:dialects});
  \item automatic sparse differentiation that preserves repeated structure, so that the derivative of a function applied at every stage is again a single function applied at every stage, which computes only the nonzero entries of the derivative (Section~\ref{sec:sparse-ad});
  \item a sequence of optimizations that exploit both the repeated structure of the problem and the properties of the target processor, such as expanding each stage into scalar code inside a loop and vectorizing that loop across stages (Section~\ref{sec:passes}).
\end{itemize}
\scaly{} is open source\footnote{\url{https://github.com/PREDICT-EPFL/scaly}} and written in pure Python, generates dependency-free C code with a \casadi{}-compatible calling convention, and provides interfaces to existing optimization solvers such as IPOPT~\cite{wachter2006ipopt} and PIQP~\cite{schwan2023piqp}.
Section~\ref{sec:background} lists what makes generated code fast and reviews related tools, Section~\ref{sec:method} follows a running example through the compiler, and Section~\ref{sec:experiments} compares \scaly{} with \casadi{}.

\section{Background and Related Work}\label{sec:background}

\begin{table*}[t]
  \centering
  \caption{Tools that compile numerical functions from a symbolic or traced description.
  The \casadi{} row covers its \texttt{SX} and \texttt{MX} graphs (Section~\ref{sec:tools}) and \texttt{Function.map} (Section~\ref{sec:exp-setup}), between which the user chooses by hand.}
  \label{tab:tools}
  \footnotesize
  \setlength{\tabcolsep}{5pt}
  \newcommand{\hd}[1]{\begin{tabular}[b]{@{}l@{}}#1\end{tabular}}
  \begin{tabular}{@{}lllllll@{}}
    \toprule
    Tool & \hd{Execution\\from Python} & \hd{Exported\\artifact} & \hd{Runtime of\\exported artifact} & \hd{Sparse\\derivatives} & \hd{Machine code\\independent of $N$} & \hd{Scalar code\\or loops} \\
    \midrule
    \casadi{}~\cite{andersson2019casadi} & interpreter or compiled C & C source & none & yes & with \texttt{map} & chosen by hand \\
    SymForce~\cite{martiros2022symforce} & generated Python & C++ source & Eigen & yes & no & scalar only \\
    JAX~\cite{jax} & compiled by XLA & StableHLO & XLA & no & with \texttt{vmap} or \texttt{scan} & chosen by compiler \\
    PyTorch~\cite{ansel2024pytorch2} & eager or \texttt{torch.compile} & AOTInductor library & libtorch & no & with \texttt{vmap} & chosen by compiler \\
    IREE~\cite{iree} & IREE runtime & VM bytecode & IREE runtime & no AD & depends on input & chosen by compiler \\
    \midrule
    \scaly{} & compiled C & C source & none & yes & yes & chosen per function \\
    \bottomrule
  \end{tabular}
\end{table*}

\subsection{Fast Code on Modern CPUs}\label{sec:fast-code}
Three largely independent principles make numerical C code fast on modern processors~\cite{hennessy2017architecture}.
\begin{itemize}
  \item \emph{Memory access}: a processor computes far faster than it reads main memory, which it fetches in small blocks into a cache, so code is fast when the values it reads lie next to each other in memory and when an intermediate result is used right after it is computed, so that it stays in a register instead of being written to an array and read back.
  \item \emph{Unrolling}: unrolling a loop over array entries gives \emph{scalar code}, one straight-line statement per entry, which the C compiler can simplify with the constants it sees, for instance by removing a multiplication by an entry that is always zero.
  With constant coefficients $a = (2, 0, 3)$, for example, unrolling the loop on the left gives the scalar code on the right, in which the multiplication by zero has disappeared:
\par\smallskip{\centering\scriptsize\begin{tabular}{l|l}
  Loop & Scalar code \\ \hline
  \texttt{\textbf{for} (k = 0; k < 3; ++k)} & \texttt{y[0] = 2*x[0];} \\
  \texttt{\hspace{1em}y[k] = a[k]*x[k];} & \texttt{y[1] = 0;} \\
  & \texttt{y[2] = 3*x[2];}
\end{tabular}\par}\smallskip
  Scalar code is therefore faster to evaluate, but its source and machine code grow with the number of iterations, so that a long scalar function compiles slowly and eventually not at all.
  \item \emph{Vectorization}: single instruction, multiple data (SIMD) instructions apply one operation to all the values in a vector register, typically two to eight double-precision numbers.
  A loop can use them when its iterations repeat the same operations on consecutive, independent data.
  C compilers vectorize a loop automatically only when they can prove that this preserves its result, so the C code should make this visible, with contiguous data and independent iterations.
  A loop that calls a math function such as $\sin$ can only be vectorized if a vector math library provides a version of that function that processes a whole vector register at once.
\end{itemize}

\subsection{Related Tools}\label{sec:tools}
Existing tools make different trade-offs between the speed of the generated code, its growth with the size of the problem, and its dependencies, and Table~\ref{tab:tools} compares them along these lines.
\casadi{}~\cite{andersson2019casadi} obtains sparse derivatives by propagating sparsity patterns through its graph of operations and coloring the result (Section~\ref{sec:sparse-ad}).
It builds this graph from one of two symbolic types, \texttt{SX} for scalar expressions and \texttt{MX} for matrix expressions, and writes it out as C code one operation at a time, so the type the user chooses by hand fixes the form of the code.
Scalar expressions built with \texttt{SX} give scalar code, which exposes every zero entry and is excellent for small problems but unrolls repeated functions, such as the stages of a horizon, so that the code grows with the number of repetitions, while matrix expressions built with \texttt{MX} give compact code that evaluates each operation as a whole.
\casadi{} simplifies expressions, eliminates common subexpressions, and reuses temporary storage, but it does not fuse loops, arrange memory, or vectorize, which it leaves to the C compiler.
SymForce~\cite{martiros2022symforce} generates code from scalar symbolic expressions, like \texttt{SX}, and therefore shares its growth with the horizon.
Machine-learning compilers, such as XLA~\cite{xla} behind JAX~\cite{jax}, TorchInductor behind PyTorch~\cite{ansel2024pytorch2}, and IREE~\cite{iree}, target dense tensor operations, for which loop optimizations are well understood~\cite{ragankelley2013halide}.
Their AD does not exploit the fixed sparsity patterns of Jacobians and Hessians.
The sparse tensor support of MLIR~\cite{mlir2020, bik2022mlirsparse} handles sparsity given as data at runtime, which the generated code must traverse through index arrays, rather than the fixed patterns of an OCP, which a compiler can specialize on.
It is also not integrated into end-to-end compilers such as XLA and IREE.

\casadi{} is the only one of these tools that covers the same workflow as \scaly{}, from a Python description to sparse derivatives, solver interfaces, and C code that is both compiled for use from Python and exported for deployment, so the experiments compare \scaly{} with \casadi{}.
Benchmarking JAX or PyTorch on our problems would require adding to them sparse differentiation that preserves the problem structure, which neither currently provides, and exporting their compiled code ahead of time (AOT) through paths that deviate from the just-in-time (JIT) pipelines these tools are designed around.
Such a comparison would measure our additions rather than the tools their users run, so we leave it out.

\section{Compiler Design}\label{sec:method}

\scaly{} offers a modeling language embedded in Python together with an optimizing compiler that turns mathematical functions, such as the cost, constraints, and derivatives of a nonlinear program (NLP), into C.
A user writes each function in Python with named inputs and outputs of fixed shape, and its derivatives are again functions, which can be called, differentiated further, and compiled like any other.
Like most compilers~\cite{aho2006compilers}, \scaly{} stores a function in an \emph{intermediate representation} (IR), a graph whose nodes are operations and whose edges carry the result of one operation to the next.
A \emph{pass} rewrites this graph into an equivalent one that is simpler or faster to evaluate, and \scaly{} compiles a function through a sequence of passes.
\scaly{} uses two IRs, called \emph{dialects} as in MLIR~\cite{mlir2020}, and a step called \emph{lowering} translates a function from the first into the second.

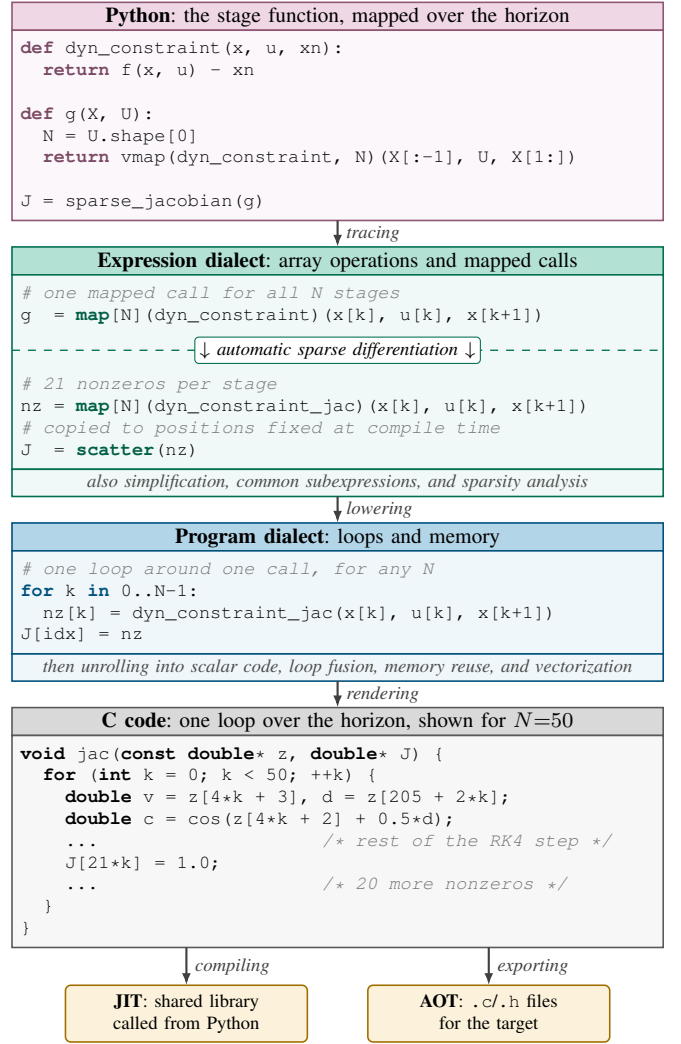
\begin{figure}[!t]
  \centering
\lstdefinestyle{ov}{language={}, basicstyle=\ttfamily\fontsize{7}{8.2}\selectfont, frame=none, aboveskip=0pt, belowskip=0pt,
  keywordstyle=\bfseries, columns=fullflexible, keepspaces=true, commentstyle=\itshape\color{black!55}, morecomment=[l]{\#}}
\lstdefinestyle{ov-py}{style=ov, language=Python, keywordstyle=\bfseries\color{c-mapsx!60!black}}
\lstdefinestyle{ov-expr}{style=ov, keywordstyle=\bfseries\color{c-scaly!60!black}, morekeywords={map,scatter}}
\lstdefinestyle{ov-prog}{style=ov, keywordstyle=\bfseries\color{c-mx!70!black}, morekeywords={for,in}}
\lstdefinestyle{ov-c}{style=ov, language=C, keywordstyle=\bfseries, morecomment=[s]{/*}{*/}}
\newlength{\ovw}\setlength{\ovw}{\dimexpr\columnwidth-7pt\relax}
\newsavebox{\ovpy}\newsavebox{\ovexpr}\newsavebox{\ovjac}\newsavebox{\ovprog}\newsavebox{\ovc}
\begin{lrbox}{\ovpy}\begin{minipage}{\ovw}
\begin{lstlisting}[style=ov-py]
def dyn_constraint(x, u, xn):
  return f(x, u) - xn

def g(X, U):
  N = U.shape[0]
  return vmap(dyn_constraint, N)(X[:-1], U, X[1:])

J = sparse_jacobian(g)
\end{lstlisting}
\end{minipage}\end{lrbox}
\begin{lrbox}{\ovexpr}\begin{minipage}{\ovw}
\begin{lstlisting}[style=ov-expr]
# one mapped call for all N stages
g  = map[N](dyn_constraint)(x[k], u[k], x[k+1])
\end{lstlisting}\vspace{6pt}
\end{minipage}\end{lrbox}
\begin{lrbox}{\ovjac}\begin{minipage}{\ovw}\vspace{6pt}
\begin{lstlisting}[style=ov-expr]
# 21 nonzeros per stage
nz = map[N](dyn_constraint_jac)(x[k], u[k], x[k+1])
# copied to positions fixed at compile time
J  = scatter(nz)
\end{lstlisting}
\end{minipage}\end{lrbox}
\begin{lrbox}{\ovprog}\begin{minipage}{\ovw}
\begin{lstlisting}[style=ov-prog]
# one loop around one call, for any N
for k in 0..N-1:
  nz[k] = dyn_constraint_jac(x[k], u[k], x[k+1])
J[idx] = nz
\end{lstlisting}
\end{minipage}\end{lrbox}
\begin{lrbox}{\ovc}\begin{minipage}{\ovw}
\begin{lstlisting}[style=ov-c]
void jac(const double* z, double* J) {
  for (int k = 0; k < 50; ++k) {
    double v = z[4*k + 3], d = z[205 + 2*k];
    double c = cos(z[4*k + 2] + 0.5*d);
    ...                    /* rest of the RK4 step */
    J[21*k] = 1.0;
    ...                    /* 20 more nonzeros */
  }
}
\end{lstlisting}
\end{minipage}\end{lrbox}
\begin{tikzpicture}[
  font=\scriptsize, node distance=0pt, outer sep=0pt,
  head/.style={draw=#1!70!black, fill=#1!30, line width=0.6pt, minimum width=\columnwidth-0.6pt, inner sep=1.5pt,
    font=\footnotesize, text height=6pt, text depth=1.5pt},
  code/.style={draw=#1!70!black, fill=#1!6, line width=0.6pt, minimum width=\columnwidth-0.6pt, inner sep=3pt},
  tags/.style={draw=#1!70!black, fill=#1!6, line width=0.6pt, text width=\columnwidth-4.6pt, align=center, inner sep=2pt,
    font=\scriptsize\itshape, text=black!70},
  stage/.style={draw=#1!70!black, fill=#1!14, rounded corners=2pt, align=center, inner sep=2.5pt, line width=0.6pt},
  arrow/.style={-{Latex[length=1.6mm]}, thick, black!75},
  note/.style={right, font=\scriptsize\itshape, black!75},
  gap/.style={below=3.5mm of #1},
]
\node[head=c-mapsx] (hpy) {\textbf{Python}: the stage function, mapped over the horizon};
\node[code=c-mapsx, below=of hpy] (py) {\usebox{\ovpy}};
\node[head=c-scaly, gap=py] (hexpr) {\textbf{Expression dialect}: array operations and mapped calls};
\node[code=c-scaly, draw=none, below=of hexpr] (expr) {\usebox{\ovexpr}};
\node[code=c-scaly, draw=none, below=of expr] (jac) {\usebox{\ovjac}};
\draw[c-scaly!70!black, line width=0.6pt] (expr.north west) rectangle (jac.south east);
\node[tags=c-scaly, below=of jac] (etags) {also simplification, common subexpressions, and sparsity analysis};
\node[head=c-mx, gap=etags] (hprog) {\textbf{Program dialect}: loops and memory};
\node[code=c-mx, below=of hprog] (prog) {\usebox{\ovprog}};
\node[tags=c-mx, below=of prog] (ptags) {then unrolling into scalar code, loop fusion, memory reuse, and vectorization};
\node[head=gray, gap=ptags] (hc) {\textbf{C code}: one loop over the horizon, shown for $N{=}50$};
\node[code=gray, below=of hc] (c) {\usebox{\ovc}};
\node[stage=c-callmx, minimum width=3.2cm, minimum height=0.75cm, below=5mm of c, xshift=-2cm] (jit) {\textbf{JIT}: shared library\\called from Python};
\node[stage=c-callmx, minimum width=3.2cm, minimum height=0.75cm, below=5mm of c, xshift=2cm] (aot) {\textbf{AOT}: \texttt{.c}/\texttt{.h} files\\for the target};
\draw[c-scaly!70!black, dashed, line width=0.5pt] (expr.south west) -- (expr.south east);
\node[draw=c-scaly!70!black, fill=white, rounded corners=1.5pt, inner sep=1.8pt, font=\scriptsize\itshape]
  at (expr.south) {$\downarrow$ automatic sparse differentiation $\downarrow$};
\draw[arrow] (py) -- node[note] {tracing} (hexpr);
\draw[arrow] (etags) -- node[note] {lowering} (hprog);
\draw[arrow] (ptags) -- node[note] {rendering} (hc);
\draw[arrow] (c.south -| jit) -- node[note] {compiling} (jit);
\draw[arrow] (c.south -| aot) -- node[note] {exporting} (aot);
\end{tikzpicture}
  \caption{\scaly{} on the running example, the dynamics constraints $g_k = f(x_k, u_k) - x_{k+1}$ of multiple shooting and their sparse Jacobian.
  The stage function stays a single mapped call through differentiation and becomes one loop after lowering, and the program-level optimizations expand the body of this loop into scalar statements.
  The Python code uses a simplified version of the interface of \scaly{}, the listings of the two dialects are pseudocode, and the generated C code is abridged and simplified.}
  \label{fig:overview}
\end{figure}

This section and Fig.~\ref{fig:overview} follow one function through the compiler: the dynamics constraints $g_k = f(x_k, u_k) - x_{k+1}$, $k = 0, \dots, N-1$, which arise in multiple shooting and whose Jacobian the solver evaluates at every iteration.
In this example, $f$ is one step of the fourth-order Runge--Kutta method (RK4) applied to the race-car dynamics of Section~\ref{sec:exp-setup}, with state $x_k \in \mathbb{R}^4$ and control $u_k \in \mathbb{R}^2$.
Fig.~\ref{fig:overview} shows this function at each step, with each IR printed as pseudocode, one operation per line.

\subsection{Two Dialects}\label{sec:dialects}
\scaly{} uses two dialects because its transformations need different levels of abstraction: differentiation and algebraic simplification need whole array operations, whereas unrolling, loop fusion, and memory reuse need explicit loops and buffers.
The \emph{expression dialect} (second box in Fig.~\ref{fig:overview}) encodes the mathematics of a function.
Its operations act on whole arrays of fixed shape, such as a matrix product, a sum over an axis, or an elementwise function, or call other functions.
It says nothing about loops, memory, or evaluation order.
The \emph{program dialect} (third box in Fig.~\ref{fig:overview}) encodes the execution of a function, with explicit buffers, loops, loads, stores, and calls to subroutines.
A matrix product, for instance, is one operation in the expression dialect, which differentiation handles with one rule, but after lowering it becomes nested loops that accumulate sums, from which differentiation would first have to recover the product.
Reusing memory, conversely, needs buffers, which exist only after lowering.
\scaly{} therefore performs all mathematical transformations before lowering and lowers each function exactly once.
Every value has a static shape, so all memory is known at compile time and the generated code never allocates memory dynamically.

\scaly{} builds the expression IR of a function by \emph{tracing}, the first arrow of Fig.~\ref{fig:overview}: the user's Python function runs once on symbolic inputs, and its operations are recorded.
Tracing is simple to implement and lets the user organize a function with ordinary Python functions and classes, since only the resulting operations are recorded.
Its main downside for numerical code is that it unrolls Python loops: a Python loop over $k$ that calls \texttt{dyn\_constraint} leaves $N$ copies of it in the trace.
The user therefore declares repeated structure explicitly with a \emph{mapped call}, \texttt{vmap} in Fig.~\ref{fig:overview}, which applies one function to $N$ parts of its inputs and stays a single operation, \texttt{map[N]}, in the expression dialect.
Mapped calls generalize the \texttt{vmap} of JAX: an input can be split into disjoint parts or read through overlapping windows.
The dynamics constraints use the latter, written in Fig.~\ref{fig:overview} as the two shifted slices \texttt{X[:-1]} and \texttt{X[1:]}, since the windows $(x_k, u_k, x_{k+1})$ of consecutive stages share a state.

\subsection{Expression-Level Transformations}\label{sec:sparse-ad}
At the expression level, \scaly{} simplifies algebraic expressions, precomputes constant subexpressions, eliminates common subexpressions, and differentiates.
Differentiation computes the sparse Jacobians and Hessians that the solver needs by automatic sparse differentiation (ASD)~\cite{gebremedhin2005jacobian}.
Forward-mode AD computes a Jacobian-vector product at a small multiple of the cost of the function~\cite{griewank2008evaluating}, so a dense Jacobian with $n$ columns costs $n$ products.
ASD first analyzes sparsity by propagating Boolean dependency patterns through the graph to find which entries can be nonzero.
It then groups columns that share no nonzero row by graph coloring~\cite{gebremedhin2005jacobian}, so that one product with the sum of the unit vectors of a group, called a \emph{seed}, recovers all columns of the group.

\begin{figure}[t]
  \centering
\begin{tikzpicture}[
  font=\scriptsize, x=0.45cm, y=0.45cm,
  cell/.style={draw=gray!60, line width=0.3pt},
]
\def\colors{{"c-mx","c-callmx","c-mapsx","c-mx","c-callmx","c-mx","c-mapsx","c-mx","c-callmx"}}
\def\seeds{{1,2,3,1,2,1,3,1,2}}
\def\labels{{"$x_0$","$u_0$","$x_1$","$u_1$","$x_2$","$u_2$","$x_3$","$u_3$","$x_4$"}}
\foreach \r in {0,...,3} {
  \foreach \c in {0,...,8} {
    \draw[cell, fill=white] (\c, -\r) rectangle ++(1, 1);
  }
  \foreach \j in {0,1,2} {
    \pgfmathtruncatemacro{\c}{2*\r+\j}
    \pgfmathsetmacro{\col}{\colors[\c]}
    \pgfmathtruncatemacro{\seed}{\seeds[\c]}
    \draw[cell, fill=\col!70] (\c, -\r) rectangle ++(1, 1) node[midway] {\seed};
  }
  \node[anchor=east] at (0, -\r+0.5) {$g_{\r}$};
}
\foreach \c in {0,...,8} {
  \pgfmathsetmacro{\lab}{\labels[\c]}
  \node[anchor=north] at (\c+0.5, -3) {\lab};
}
\node[anchor=south] at (4.5, 1) {$J$};
\draw[-{Latex[length=1.6mm]}, thick] (9.6, -1) -- node[above] {$S$} (11.4, -1);
\foreach \r in {0,...,3} {
  \foreach \j/\col in {0/c-mx, 1/c-callmx, 2/c-mapsx} {
    \draw[cell, fill=\col!70] (12+\j, -\r) rectangle ++(1, 1) node[midway] {\the\numexpr\j+1\relax};
  }
}
\node[anchor=south] at (13.5, 1) {$JS$};
\end{tikzpicture}
  \caption{Sparse Jacobian $J$ of the dynamics constraints $g_k = f(x_k, u_k) - x_{k+1}$ for $N{=}4$, with a scalar state and control instead of the vehicle of Fig.~\ref{fig:overview}.
  Columns of the same color and number share no nonzero row, so multiplying $J$ by the matrix $S$ whose three columns are the seeds, each the sum of the unit vectors of one color, gives three Jacobian-vector products $JS$ that together contain every nonzero of $J$.
  \scaly{} obtains these nonzeros from one mapped call of a stage derivative that returns the three nonzeros of each row, for any $N$.}
  \label{fig:defect-jacobian}
\end{figure}
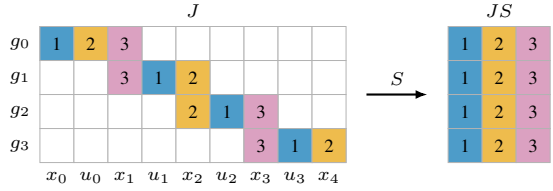

\scaly{} differentiates a mapped call into a mapped call of the derivative of the mapped function, as below the dashed line in Fig.~\ref{fig:overview}, so that differentiation does not copy the function along the horizon.
Fig.~\ref{fig:defect-jacobian} shows the Jacobian of the dynamics constraints for a scalar state and control.
Each constraint $g_k$ depends only on $x_k$, $u_k$, and $x_{k+1}$, so three colors suffice for any horizon, but the colors that stage $k$ sees change from one stage to the next.
Passing these stage-dependent seeds as runtime inputs, as in the mapped \texttt{SX} of \casadi{} (Section~\ref{sec:exp-setup}), would keep products with zeros and ones from simplifying away.
\scaly{} therefore colors the Jacobian of the stage function itself, with respect to each of its arguments, and substitutes the resulting seeds into the stage derivative as constants.
Products with zeros and ones then simplify away, and the stage derivative returns only the three nonzeros $\partial f / \partial x_k$, $\partial f / \partial u_k$, and $-1$ of row $k$.
One mapped call of this derivative computes the nonzeros of all stages in a single loop, and \texttt{scatter} in Fig.~\ref{fig:overview} copies them to positions in the sparse Jacobian that are fixed at compile time.
Reverse-mode AD computes the gradient at a small multiple of the cost of the function, whatever the number of inputs~\cite{griewank2008evaluating}, while preserving mapped calls in the same way.
The Hessian is the Jacobian of the gradient, obtained in forward mode with a coloring that exploits its symmetry and can therefore use fewer seeds~\cite{gebremedhin2005jacobian}.

When stages are coupled only to their neighbors, as here, $N$ appears in the generated code only as a loop bound, in index arithmetic, and in the length of constant sparsity and index tables.
The machine code and the build time therefore stay nearly constant in $N$, while the constant tables and the evaluation time grow linearly with $N$.

\subsection{Lowering and Program-Level Optimizations}\label{sec:passes}
Lowering translates each array operation of the expression dialect into loops over buffers, and a mapped call into a loop of $N$ iterations around one call to the mapped function, so that the generated code contains this function once, for any $N$.
In Fig.~\ref{fig:overview}, the mapped stage derivative of the running example thus becomes the loop of the program-dialect box.
\scaly{} then applies the three principles of Section~\ref{sec:fast-code} to the program dialect, where it knows more about the code than a C compiler can recover, for instance that the $N$ iterations of a loop apply one function and which entries of a derivative are always zero.
\begin{itemize}
  \item \emph{Unrolling}: \scaly{} decides for each scope whether to unroll its loops into scalar code, which it does by default when their bounds are constant and the scalar code stays within a size limit that trades off the speed of scalar code against code growth and compile time (Section~\ref{sec:fast-code}).
  A small stage, such as the stage derivative of the running example, thus becomes straight-line code inside the loop over the horizon, in which simplification removes every multiplication by an entry that is always zero, while a large neural network layer exceeds the limit and remains a loop, which costs little since its dense weights leave little to simplify.
  \item \emph{Memory access}: loop fusion merges consecutive loops over the same elements, so that the values passed from one to the next stay in registers instead of in an intermediate array, unless fusion would duplicate an expensive operation such as an exponential, the same trade-off as in Halide~\cite{ragankelley2013halide}.
  Temporaries that are never needed at the same time share one buffer, which keeps the working memory of a function small, so that more of it fits in cache.
  \item \emph{Vectorization}: \scaly{} vectorizes the loop over stages rather than leaving it to the C compiler, since the iterations of a mapped call are independent by construction, so that vectorization does not depend on what the C compiler can prove from index arithmetic and pointers.
  It uses explicit vector types that let each iteration process as many consecutive stages as fit in a vector register of the target, and calls the vectorized math functions of a vector math library when the target provides one.
\end{itemize}

\subsection{Generated Code}\label{sec:generated}
Rendering translates the optimized program into C almost line by line, since the program dialect already fixes its loops, buffers, and statements.
For the Jacobian of the running example, at the bottom of Fig.~\ref{fig:overview}, the loop body evaluates the RK4 step and its derivative once per stage, computing each trigonometric function they share only once, and writes the 21 nonzeros of that stage.
It stores constant entries directly, such as the 1.0 in Fig.~\ref{fig:overview}, the derivative of the next position with respect to the current one, which is constant because the dynamics do not depend on the position.
The generated C code follows the calling convention of \casadi{}, so that tools built on the code generated by \casadi{}, such as acados, can switch to the code generated by \scaly{} without any changes.
It depends only on the C math library, whose vectorized functions it calls when they are available (Section~\ref{sec:passes}).
For an NLP, \scaly{} generates the objective, the constraints, their first derivatives, and the Hessian of the Lagrangian, together with a wrapper around IPOPT~\cite{wachter2006ipopt}, PIQP~\cite{schwan2023piqp}, or the sequential quadratic programming (SQP) solver included in \scaly{}, so that a whole solve runs in C.
The same C code is either compiled JIT into a shared library called from Python or exported AOT as sources and headers for the build system of the target, so the code tested on a workstation is the code deployed.
By default, JIT code targets the processor that runs it, with the width of its vector registers and its vector math library, while AOT code can target the same features or leave them out to stay portable across processors.

\begin{figure*}[t]
  \centering
\begin{tikzpicture}
\begin{groupplot}[
  bench, xmode=log, ymode=log,
  group style={group size=3 by 3, horizontal sep=0.95cm, vertical sep=0.35cm,
    xticklabels at=edge bottom, xlabels at=edge bottom, ylabels at=edge left},
  width=0.33\textwidth, height=3.2cm,
]
\nextgroupplot[title={Race-car MPC}, ylabel={Eval.\ [\si{\micro\second}]}, xmin=4, xmax=1200, xtick={10,100},
  legend to name=benchlegend, legend columns=4,
  legend cell align=left, legend style={font=\footnotesize, draw=none, /tikz/every even column/.append style={column sep=0.8em}}]
\pgfplotsinvokeforeach{scaly,casadi_sx,casadi_mx,casadi_call_mx,casadi_map_sx,casadi_mx_gemm,best}{\addlegendimage{bk/#1}}
\addlegendimage{only marks, mark=*, mark options={fill=white}, black!60}
\legend{Scaly, CasADi SX, CasADi MX, MX + stage calls, mapped SX, MX + batched layers, best CasADi, stopped at \SI{180}{\second}}
\benchseries{data/hessian_race_cars.csv}{rt}
\benchratio{data/hessian_race_cars.csv}{rt}{6}{anchor=west}
\nextgroupplot[title={Neural-process MPC}, xmin=5, xmax=500]
\benchseries{data/hessian_npmpc.csv}{rt}
\benchratio{data/hessian_npmpc.csv}{rt}{5}{anchor=west}
\nextgroupplot[title={Unbumpercars filter}, xmin=1.7, xmax=38, xtick={2,4,8,16,32}]
\benchseries{data/hessian_unbumpercars.csv}{rt}
\benchratio{data/hessian_unbumpercars.csv}{rt}{2}{anchor=west}
\nextgroupplot[ylabel={Build [\si{\second}]}, xmin=4, xmax=1200, xtick={10,100}, ymin=0.3, ymax=600]
\benchseries{data/hessian_race_cars.csv}{bt}
\timeouts{data/hessian_race_cars.csv}
\benchratio{data/hessian_race_cars.csv}{bt}{6}{anchor=west}
\draw[gray, dashed] ({rel axis cs:0,0}|-{axis cs:5,180}) -- ({rel axis cs:1,0}|-{axis cs:5,180});
\nextgroupplot[xmin=5, xmax=500, ymin=0.3, ymax=600]
\benchseries{data/hessian_npmpc.csv}{bt}
\timeouts{data/hessian_npmpc.csv}
\benchratio{data/hessian_npmpc.csv}{bt}{5}{anchor=west}
\draw[gray, dashed] ({rel axis cs:0,0}|-{axis cs:6,180}) -- ({rel axis cs:1,0}|-{axis cs:6,180});
\nextgroupplot[xmin=1.7, xmax=38, xtick={2,4,8,16,32}, ymin=0.3, ymax=600]
\benchseries{data/hessian_unbumpercars.csv}{bt}
\timeouts{data/hessian_unbumpercars.csv}
\benchratio{data/hessian_unbumpercars.csv}{bt}{2}{anchor=west}
\draw[gray, dashed] ({rel axis cs:0,0}|-{axis cs:2,180}) -- ({rel axis cs:1,0}|-{axis cs:2,180});
\nextgroupplot[ylabel={Binary [\si{\kilo\byte}]}, xlabel={horizon $N$}, xmin=4, xmax=1200, xtick={10,100}]
\benchseries{data/hessian_race_cars.csv}{os}
\benchratio{data/hessian_race_cars.csv}{os}{6}{anchor=west}
\nextgroupplot[xlabel={horizon $N$}, xmin=5, xmax=500]
\benchseries{data/hessian_npmpc.csv}{os}
\benchratio{data/hessian_npmpc.csv}{os}{5}{anchor=west}
\nextgroupplot[xlabel={cars $C$}, xmin=1.7, xmax=38, xtick={2,4,8,16,32}]
\benchseries{data/hessian_unbumpercars.csv}{os}
\benchratio{data/hessian_unbumpercars.csv}{os}{2}{anchor=west}
\end{groupplot}
\node[anchor=south] (legend) at ($(group c1r1.north west)!0.5!(group c3r1.north east) + (0,0.45cm)$) {\pgfplotslegendfromname{benchlegend}};
\coordinate (bbw) at (current bounding box.west);
\coordinate (bbe) at (current bounding box.east);
\pgfresetboundingbox
\path (bbw |- group c1r3.outer south) rectangle (bbe |- legend.north);
\end{tikzpicture}
  \caption{Evaluation time (top), build time from the problem description to the object file with the \SI{180}{\second} compile limit dashed (middle), and binary size (bottom) of the Hessian of the Lagrangian as each problem grows.
  The gray line follows the best \casadi{} implementation at each size, the arrows give the ratio of the best \casadi{} value to that of \scaly{} at the largest size where \casadi{} compiles, and every marker above the dashed limit marks a compile that was stopped.}
  \label{fig:scalability}
\end{figure*}
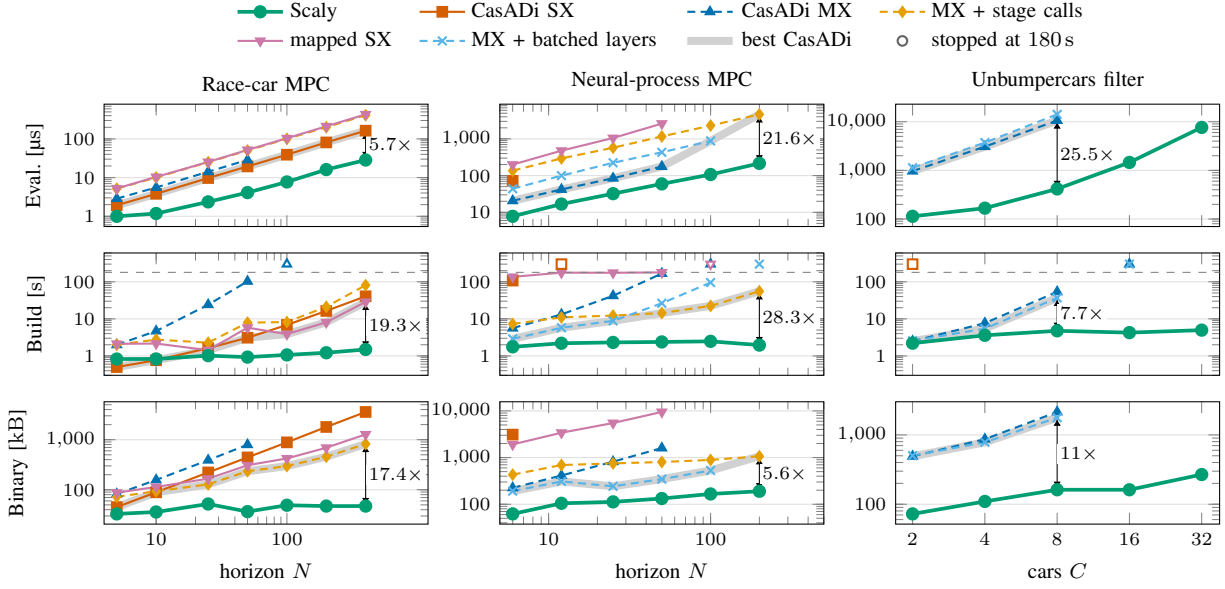

\section{Experiments}\label{sec:experiments}
We evaluate the functions generated by \scaly{}~0.1.0 against those of \casadi{}~3.8.0, first by how their evaluation time, build time, and binary size scale with the size of the problem (Section~\ref{sec:exp-scalability}), then by the solver time per control step of complete controllers in closed-loop simulations (Section~\ref{sec:exp-closed-loop}).
The three problems cover first-principles and neural models, horizons and sets of agents, and solves in which function evaluation takes from under a tenth to nearly all of the time.

\begin{figure*}[t]
  \centering
\newcommand{\clpanel}[2]{%
  \addplot[bar, rows={#1}{#2}, fill=c-scaly!30] table[x=total_scaly, y expr=\ys] {data/closed_loop.csv};
  \addplot[bar, rows={#1}{#2}, fill=c-scaly] table[x=fe_scaly, y expr=\ys] {data/closed_loop.csv};
  \addplot[bar, rows={#1}{#2}, fill=c-casadi!30] table[x=total_casadi, y expr=\yc] {data/closed_loop.csv};
  \addplot[bar, rows={#1}{#2}, fill=c-casadi] table[x=fe_casadi, y expr=\yc] {data/closed_loop.csv};
  \addplot[rows={#1}{#2}, only marks, mark size=0pt, nodes near coords, nodes near coords align={anchor=west},
    every node near coord/.append style={font=\scriptsize, inner sep=1pt, xshift=1pt},
    point meta=\thisrow{speedup_total}, visualization depends on={\thisrow{speedup_fe}\as\spfe},
    nodes near coords={$\pgfmathprintnumber[fixed, zerofill, precision=2]{\pgfplotspointmeta}\times$
      (FE $\pgfmathprintnumber[fixed, zerofill, precision=2]{\spfe}\times$)}]
    table[x=fe_scaly, y expr=\ys] {data/closed_loop.csv};
}
\def\ys{2*mod(\coordindex,2)}
\def\yc{2*mod(\coordindex,2) + 1}
\begin{tikzpicture}
\begin{groupplot}[
  bench, ymajorgrids=false, clip=false, xtick pos=lower,
  group style={group size=3 by 1, horizontal sep=0.55cm, yticklabels at=edge left},
  scale only axis, width=4.8cm, height=1.3cm,
  xmin=0, enlarge x limits={upper, value=0.05},
  y dir=reverse, ymin=-0.6, ymax=3.6, ytick={0,1,2,3},
  yticklabels={{SQP, \scaly{}}, {SQP, \casadi{}}, {IPOPT, \scaly{}}, {IPOPT, \casadi{}}},
  ytick style={draw=none},
  table/col sep=comma,
  bar/.style={xbar, bar width=0.75, draw=none, area legend, forget plot},
  rows/.style 2 args={skip coords between index={0}{#1}, skip coords between index={#2}{6}},
]
\nextgroupplot[title={Race-car MPC}, legend to name=cllegend, legend columns=4,
  legend image post style={draw=none}, legend style={font=\footnotesize, draw=none,
    /tikz/every even column/.append style={column sep=0.6em}}]
\addlegendimage{area legend, draw=none, fill=c-scaly}\addlegendentry{\scaly{}}
\addlegendimage{area legend, draw=none, fill=c-casadi}\addlegendentry{\casadi{}}
\addlegendimage{area legend, draw=none, fill=black!60}\addlegendentry{function evaluation}
\addlegendimage{area legend, draw=none, fill=black!15}\addlegendentry{rest of the solver}
\clpanel{0}{2}
\nextgroupplot[title={Neural-process MPC}]
\clpanel{2}{4}
\nextgroupplot[title={Unbumpercars filter}, xtick distance=50]
\clpanel{4}{6}
\end{groupplot}
\node[anchor=north, font=\footnotesize] at ($(group c1r1.south west)!0.5!(group c3r1.south east) + (0,-0.4cm)$) {Mean solver time per step [\si{\milli\second}]};
\node[anchor=south] at ($(group c1r1.north west)!0.5!(group c3r1.north east) + (0,0.3cm)$) {\pgfplotslegendfromname{cllegend}};
\end{tikzpicture}
  \caption{Mean solver time per control step with the same solver and functions generated by \scaly{} or \casadi{}, split into function evaluation (FE, dark) and the rest of the solver (light).
  The labels give the speedup of \scaly{} for the whole step and, in parentheses, for function evaluation.}
  \label{fig:closed-loop}
\end{figure*}
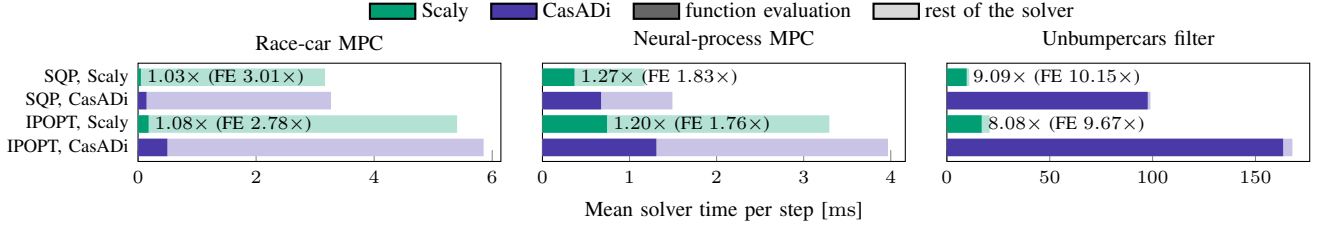

\subsection{Problems and Setup}\label{sec:exp-setup}
The \emph{race-car MPC} tracks a reference along a race track over a horizon of $N$ stages with a kinematic bicycle model~\cite{kong2015kinematic}.
Its state $x = (p_x, p_y, \theta, v)$ holds the position, heading, and speed of the car, its control $u = (T, \delta)$ holds the throttle and the steering angle, and its dynamics
\begin{align*}
  \dot p_x &= v \cos(\theta + \beta), & \dot p_y &= v \sin(\theta + \beta), \\
  \dot \theta &= \tfrac{2v}{\ell} \sin\beta, & \dot v &= \tfrac{1}{m} \big(C_m T - F_r(v \cos\beta)\big),
\end{align*}
with slip angle $\beta = \delta / 2$, wheelbase $\ell$, mass $m$, motor constant $C_m$, and rolling and aerodynamic resistance $F_r(v_x) = (C_{r0} + C_{r1} v_x + C_{r2} v_x^2) \tanh(10\, v_x)$, are integrated by one RK4 step of \SI{50}{\milli\second} per stage.
The cost penalizes the deviation from the reference and the control effort, and a corridor constraint keeps the front corners of the car on the track.
The \emph{neural-process MPC}~\cite{waibel2026npmpc} swings up a Furuta pendulum whose dynamics are a neural network ($9{\to}32{\to}32{\to}2$, sigmoid) integrated by the trapezoidal rule, over $N$ stages.
The \emph{unbumpercars filter} is a safety filter that keeps $C$ cars apart, with a neural car model ($6{\to}256{\to}128{\to}3$, ReLU smoothed for the filter)~\cite{petri2025sysid}, RK4 pose integration, and a barrier constraint for every pair of cars, which \scaly{} maps over cars and pairs like stages.

\casadi{} offers several ways to implement each problem, and the fastest one depends on the problem and its size.
We therefore consider five variants: \texttt{SX}, \texttt{MX} (Section~\ref{sec:tools}), mapped \texttt{SX}, \texttt{MX} with stage calls, and, on the problems with neural networks, \texttt{MX} with batched layers.
In \texttt{MX} with stage calls, each stage function remains a separate call instead of being expanded into its caller, and in \texttt{MX} with batched layers, the layers of all stages or cars are evaluated as matrix-matrix products.
In mapped \texttt{SX}, \texttt{Function.map} evaluates one \texttt{SX} stage in a loop inside an \texttt{MX} function, which keeps the code compact while simplifying each stage, but its derivative receives the seed directions as runtime inputs, so every stage multiplies by every direction, including those that are zero for that stage.
Both sides use the same compiler and flags, Clang~20.1.8 at \texttt{-O3} for the Hessians and at \texttt{-O2} for the controllers, with \texttt{-march=native} and the vector math library of glibc~2.39, as \scaly{} does by default for JIT code (Section~\ref{sec:generated}).
Every timed Hessian is first checked against an independent reference.
Each benchmark runs once, in a fresh process with an empty compilation cache, under Ubuntu~24.04 on an AMD Ryzen~9~7940HS with frequency boost disabled to reduce timing variation.
The AVX-512 instructions of this processor let \scaly{} process eight stages per iteration of a vectorized loop.
Each Hessian evaluation is repeated for at least half a second of cumulative time, so that the mean evaluation time is statistically meaningful even for evaluations that take a few microseconds.
The build time covers the whole path from the problem description to the object file: constructing and differentiating the functions, generating C, which includes the optimization passes of \scaly{}, and compiling it.
Compilation is stopped after \SI{180}{\second}, beyond which rebuilding after each edit no longer fits an interactive workflow, and an implementation that times out is not tried at larger sizes, so a missing \casadi{} point means only that no tested implementation compiled in time.
The binary size is that of the compiled object file, machine code and constant data together.
The benchmark code and a description of each problem are available at \url{https://github.com/PREDICT-EPFL/scaly/tree/28c5d9e9/benchmarks}.

\subsection{Scalability of Derivative Evaluation}\label{sec:exp-scalability}
Evaluation time limits how often a controller can solve its problem, while build time and binary size decide whether the code can be built and deployed at all (Fig.~\ref{fig:scalability}).
Against the fastest \casadi{} implementation that compiles, \scaly{} evaluates the Hessian of the Lagrangian faster at every size.
On the race car, this implementation is \texttt{SX}.
\scaly{} expands the stage into the same kind of scalar code but keeps it in a loop vectorized across stages (Section~\ref{sec:passes}), and is 1.9 times faster at $N{=}5$ and 4.7 to 5.7 times faster from $N{=}50$ on.
On the problems with neural networks, expanding each dense layer into scalar operations keeps \texttt{SX} from compiling beyond the smallest size.
\scaly{} keeps the layers as loops, like \texttt{MX}, and is 2.5 to 3.0 times faster than \texttt{MX} on the neural-process MPC up to $N{=}50$.
Beyond $N{=}50$, \texttt{MX} no longer compiles, and \scaly{} is 8.2 and 22 times faster than the best remaining implementation at $N{=}100$ and $N{=}200$.
On unbumpercars, the speedup grows from 8.5 at $C{=}2$ to 26 at $C{=}8$, and with 16 or 32 cars only \scaly{} compiles.

The \casadi{} build times grow with the horizon until they reach the compile limit, because the C sources of all its implementations grow with it, to between 2.3 and \SI{8.2}{\mega\byte} at $N{=}400$, whereas the code of \scaly{} contains each mapped function once.
The build time of \scaly{} therefore stays between 0.8 and \SI{1.5}{\second} on the race car and at most \SI{2.5}{\second} on the neural-process MPC at every horizon, 19 and 28 times shorter than that of the fastest \casadi{} build at the largest horizon.
On unbumpercars, it grows only slowly with the number of cars.
The binary of \scaly{} is smaller than that of every \casadi{} implementation at every size, by a factor of up to 17 on the race car at $N{=}400$, where \texttt{SX} reaches \SI{3.6}{\mega\byte}.
On the neural-process MPC, its binary stays between 62 and \SI{190}{\kilo\byte} across the tested horizons.

\subsection{Closed-Loop Control}\label{sec:exp-closed-loop}
Fig.~\ref{fig:closed-loop} reports the solver time per control step of complete controllers over a simulated episode, with one NLP per control step and the same solver on both sides.
The solver is either IPOPT~3.14.19 or the SQP solver of \scaly{}, which uses an exact, regularized Hessian and a filter line search modeled on that of laOPT~\cite{waibel2026laopt} and solves its subproblems with PIQP.
The race car runs at $N{=}50$ with a \SI{50}{\milli\second} sampling period, the neural-process MPC at $N{=}12$ with \SI{20}{\milli\second}, and unbumpercars with $C{=}8$ cars and \SI{100}{\milli\second}.
For every function, we use the fastest implementation of the Hessian that \casadi{} provides at these sizes, \texttt{SX} for the race car and \texttt{MX} for the two problems with neural networks.
The settings, bounds, warm starts, and tolerances are identical, and both sides take the same iterations at every step except for 1 of the 200 steps of unbumpercars with SQP, whose trajectories still agree to within $4 \times 10^{-8}$.
The timers exclude Python, the plant simulation, and building the controller.

Only function evaluation differs between the two sides, so the speedup of a whole step depends on both the speedup of function evaluation and its share of the \casadi{} step.
On the race car, function evaluation is 2.8--3.0 times faster but takes under \SI{10}{\percent} of the \casadi{} step, so the whole step is only 3--\SI{8}{\percent} faster.
On the neural-process MPC, function evaluation is 1.8 times faster and takes 33--\SI{45}{\percent} of the \casadi{} step, so the whole step is 1.2--1.3 times faster.
On unbumpercars, function evaluation is 9.7--10 times faster and takes 97--\SI{99}{\percent} of the \casadi{} step, so the whole step is 8.1--9.1 times faster.
For the neural-process MPC, whose original work~\cite{waibel2026npmpc} reduced the network to two hidden layers of 32 units to meet its sampling period, the time saved leaves room for a larger network.
Building a controller, from the problem description to compiled code, takes under \SI{3}{\second} with \scaly{} on the race car and is 2.3 and 4.5 times faster than with \casadi{} for SQP and IPOPT, respectively.
With the same two solvers, it is 3.7 and 5.0 times faster on the neural-process MPC and 5.8 and 10 times faster on unbumpercars.

\section{Conclusion}\label{sec:conclusion}
\scaly{} treats code generation for optimal control as an optimizing compiler problem, in which repeated structure survives from tracing to the generated loops and the form of the code is chosen per function.
On the benchmarks, this keeps the speed of scalar code without its growth with the horizon.
Future work will add differentiation through calls to an optimization solver, so that a learning or bilevel problem can contain an inner solve.
It will also target GPUs and batched evaluation for learning.

\section*{Acknowledgment}
We thank Roland Schwan for the insightful discussions that helped shape this work, and Simone Baratto for his help with the development of the unbumpercars benchmark.
\scaly{} has been substantially developed using AI coding agents (Claude, Codex, and others), and AI assistants helped draft and edit this manuscript, in both cases under the direction and review of the authors, who are responsible for the result.

\bibliographystyle{IEEEtran}
\bibliography{refs}

\begin{thebibliography}{10}
\providecommand{\url}[1]{#1}
\csname url@samestyle\endcsname
\providecommand{\newblock}{\relax}
\providecommand{\bibinfo}[2]{#2}
\providecommand{\BIBentrySTDinterwordspacing}{\spaceskip=0pt\relax}
\providecommand{\BIBentryALTinterwordstretchfactor}{4}
\providecommand{\BIBentryALTinterwordspacing}{\spaceskip=\fontdimen2\font plus
\BIBentryALTinterwordstretchfactor\fontdimen3\font minus
  \fontdimen4\font\relax}
\providecommand{\BIBforeignlanguage}[2]{{%
\expandafter\ifx\csname l@#1\endcsname\relax
\typeout{** WARNING: IEEEtran.bst: No hyphenation pattern has been}%
\typeout{** loaded for the language `#1'. Using the pattern for}%
\typeout{** the default language instead.}%
\else
\language=\csname l@#1\endcsname
\fi
#2}}
\providecommand{\BIBdecl}{\relax}
\BIBdecl

\bibitem{rawlings2017mpc}
J.~B. Rawlings, D.~Q. Mayne, and M.~Diehl, \emph{Model Predictive Control:
  Theory, Computation, and Design}, 2nd~ed.\hskip 1em plus 0.5em minus
  0.4em\relax Madison, WI: Nob Hill Publishing, 2017.

\bibitem{griewank2008evaluating}
A.~Griewank and A.~Walther, \emph{Evaluating Derivatives: Principles and
  Techniques of Algorithmic Differentiation}, 2nd~ed.\hskip 1em plus 0.5em
  minus 0.4em\relax Philadelphia, PA: SIAM, 2008.

\bibitem{houska2011acado}
B.~Houska, H.~J. Ferreau, and M.~Diehl, ``{ACADO} toolkit---an open-source
  framework for automatic control and dynamic optimization,'' \emph{Optimal
  Control Applications and Methods}, vol.~32, no.~3, pp. 298--312, 2011.

\bibitem{mattingley2012cvxgen}
J.~Mattingley and S.~Boyd, ``{CVXGEN}: A code generator for embedded convex
  optimization,'' \emph{Optimization and Engineering}, vol.~13, no.~1, pp.
  1--27, 2012.

\bibitem{domahidi2012forces}
A.~Domahidi, A.~U. Zgraggen, M.~N. Zeilinger, M.~Morari, and C.~N. Jones,
  ``Efficient interior point methods for multistage problems arising in
  receding horizon control,'' in \emph{Proc. 51st IEEE Conf. on Decision and
  Control (CDC)}, 2012, pp. 668--674.

\bibitem{andersson2019casadi}
J.~A.~E. Andersson, J.~Gillis, G.~Horn, J.~B. Rawlings, and M.~Diehl,
  ``{CasADi}: A software framework for nonlinear optimization and optimal
  control,'' \emph{Mathematical Programming Computation}, vol.~11, pp. 1--36,
  2019.

\bibitem{verschueren2022acados}
R.~Verschueren \emph{et~al.}, ``acados---a modular open-source framework for
  fast embedded optimal control,'' \emph{Mathematical Programming Computation},
  vol.~14, pp. 147--183, 2022.

\bibitem{salzmann2023rtn}
T.~Salzmann, E.~Kaufmann, J.~Arrizabalaga, M.~Pavone, D.~Scaramuzza, and
  M.~Ryll, ``Real-time neural-{MPC}: Deep learning model predictive control for
  quadrotors and agile robotic platforms,'' \emph{IEEE Robotics and Automation
  Letters}, 2023.

\bibitem{waibel2026npmpc}
J.~Waibel, P.~{Mello Rella}, and C.~N. Jones, ``Neural process model predictive
  control,'' \emph{European Journal of Control}, p. 101609, 2026.

\bibitem{jax}
{JAX Authors}, ``{JAX} documentation,'' \url{https://docs.jax.dev/}, 2026.

\bibitem{ansel2024pytorch2}
J.~Ansel \emph{et~al.}, ``{PyTorch} 2: Faster machine learning through dynamic
  {Python} bytecode transformation and graph compilation,'' in \emph{Proc. 29th
  ACM Int. Conf. on Architectural Support for Programming Languages and
  Operating Systems (ASPLOS)}, vol.~2, 2024, pp. 929--947.

\bibitem{wachter2006ipopt}
A.~W{\"a}chter and L.~T. Biegler, ``On the implementation of an interior-point
  filter line-search algorithm for large-scale nonlinear programming,''
  \emph{Mathematical Programming}, vol. 106, no.~1, pp. 25--57, 2006.

\bibitem{schwan2023piqp}
R.~Schwan, Y.~Jiang, D.~Kuhn, and C.~N. Jones, ``{PIQP}: A proximal
  interior-point quadratic programming solver,'' in \emph{Proc. 62nd IEEE Conf.
  on Decision and Control (CDC)}, 2023.

\bibitem{martiros2022symforce}
H.~Martiros \emph{et~al.}, ``{SymForce}: Symbolic computation and code
  generation for robotics,'' in \emph{Proc. Robotics: Science and Systems
  (RSS)}, 2022.

\bibitem{iree}
{IREE Authors}, ``{IREE}: Intermediate representation execution environment,''
  \url{https://iree.dev/}, 2026.

\bibitem{hennessy2017architecture}
J.~L. Hennessy and D.~A. Patterson, \emph{Computer Architecture: A Quantitative
  Approach}, 6th~ed.\hskip 1em plus 0.5em minus 0.4em\relax Cambridge, MA:
  Morgan Kaufmann, 2017.

\bibitem{xla}
{OpenXLA Authors}, ``{XLA}: Optimizing compiler for machine learning,''
  \url{https://openxla.org/xla}, 2026.

\bibitem{ragankelley2013halide}
J.~Ragan-Kelley, C.~Barnes, A.~Adams, S.~Paris, F.~Durand, and S.~Amarasinghe,
  ``Halide: A language and compiler for optimizing parallelism, locality, and
  recomputation in image processing pipelines,'' in \emph{Proc. 34th ACM
  SIGPLAN Conf. on Programming Language Design and Implementation (PLDI)},
  2013, pp. 519--530.

\bibitem{mlir2020}
C.~Lattner \emph{et~al.}, ``{MLIR}: A compiler infrastructure for the end of
  {Moore}'s law,'' \emph{arXiv:2002.11054}, 2020.

\bibitem{bik2022mlirsparse}
A.~Bik, P.~Koanantakool, T.~Shpeisman, N.~Vasilache, B.~Zheng, and F.~Kjolstad,
  ``Compiler support for sparse tensor computations in {MLIR},'' \emph{ACM
  Transactions on Architecture and Code Optimization}, vol.~19, no.~4, pp.
  50:1--50:25, 2022.

\bibitem{aho2006compilers}
A.~V. Aho, M.~S. Lam, R.~Sethi, and J.~D. Ullman, \emph{Compilers: Principles,
  Techniques, and Tools}, 2nd~ed.\hskip 1em plus 0.5em minus 0.4em\relax
  Addison-Wesley, 2006.

\bibitem{gebremedhin2005jacobian}
A.~H. Gebremedhin, F.~Manne, and A.~Pothen, ``What color is your jacobian?
  graph coloring for computing derivatives,'' \emph{SIAM Review}, vol.~47,
  no.~4, pp. 629--705, 2005.

\bibitem{kong2015kinematic}
J.~Kong, M.~Pfeiffer, G.~Schildbach, and F.~Borrelli, ``Kinematic and dynamic
  vehicle models for autonomous driving control design,'' in \emph{Proc. IEEE
  Intelligent Vehicles Symp. (IV)}, 2015, pp. 1094--1099.

\bibitem{petri2025sysid}
T.~Petri, S.~Baratto, and G.~Ferrari-Trecate, ``System identification for
  dynamic modeling of large steering angle vehicles,'' arXiv:2512.02803, 2025.

\bibitem{waibel2026laopt}
J.~Waibel, R.~Schwan, and C.~N. Jones, ``{laOPT}: A native {C++} optimal
  control toolbox for high-performance implementations, and application to
  racing,'' in \emph{Proc. 10th IEEE Conf. on Control Technology and
  Applications (CCTA)}, 2026.

\end{thebibliography}

\end{document}